\documentclass[letter,traditabstract,longauth]{aa}  
\usepackage{graphicx}
\usepackage{txfonts}
\begin{document}
\title{{\it Herschel} unveils a puzzling uniformity of distant dusty galaxies\thanks{Herschel is an ESA space observatory with science instruments provided by European-led Principal Investigator consortia and with important participation from NASA.}}
\author{D.~Elbaz\inst{1}
\and H.S.~Hwang\inst{1}
\and B.~Magnelli\inst{2}
\and E.~Daddi\inst{1}
\and H.~Aussel\inst{1}
\and B.~Altieri\inst{3}
\and A.~Amblard\inst{4}
\and P.~Andreani\inst{5,6}
\and V.~Arumugam\inst{7}
\and R.~Auld\inst{8}
% ...
}

\institute{Laboratoire AIM-Paris-Saclay, CEA/DSM/Irfu - CNRS - Universit\'e Paris Diderot, CE-Saclay, pt courrier 131, F-91191 Gif-sur-Yvette, France\\
 \email{delbaz@cea.fr}
\and Max-Planck-Institut f\"ur Extraterrestrische Physik (MPE), Postfach 1312, 85741, Garching, Germany
\and Herschel Science Centre, European Space Astronomy Centre, Villanueva de la Ca\~nada, 28691 Madrid, Spain
\and Dept. of Physics \& Astronomy, University of California, Irvine, CA 92697, USA
\and ESO, Karl-Schwarzschild-Str. 2, 85748 Garching bei M\"unchen, Germany
\and INAF - Osservatorio Astronomico di Trieste, via Tiepolo 11, 34143 Trieste, Italy
\and Institute for Astronomy, University of Edinburgh, Royal Observatory, Blackford Hill, Edinburgh EH9 3HJ, UK
\and Cardiff School of Physics and Astronomy, Cardiff University, Queens Buildings, The Parade, Cardiff CF24 3AA, UK
% ...
}

\date{Received March 31, 2010; accepted April, 2010}

% \abstract{}{}{}{}{} 
% 5 {} token are mandatory
 
\abstract
{The {\it Herschel} Space Observatory enables us to accurately measure the bolometric output of starburst galaxies and active galactic nuclei (AGN) by directly sampling the peak of their far-infrared (IR) emission. Here we examine whether the spectral energy distribution (SED) and dust temperature of galaxies have strongly evolved over the last 80\,\% of the age of the Universe. We discuss possible consequences for the determination of star-formation rates (SFR) and any evidence for a major change in their star-formation properties. We use {\it Herschel} deep extragalactic surveys from 100 to 500\,$\mu$m to compute total IR luminosities in galaxies down to the faintest levels, using PACS and SPIRE in the GOODS-North field (PEP and HerMES key programs). An extension to fainter luminosities is done by stacking images on 24\,$\mu$m prior positions. We show that measurements in the SPIRE bands can be used below the {\it statistical} confusion limit if information at higher spatial resolution is used, e.g. at 24\,$\mu$m, to identify "isolated" galaxies whose flux is not boosted by bright neighbors.

Below $z$$\sim$1.5, mid-IR extrapolations are correct for star-forming galaxies with a dispersion of only 40\,\% (0.15 dex), therefore similar to $z$$\sim$0 galaxies, over three decades in luminosity below the regime of ultra-luminous IR galaxies (ULIRGs, L$_{\rm IR}$$\geq$10$^{12}$ L$_{\sun}$). This narrow distribution is puzzling when considering the range of physical processes that could have affected the SED of these galaxies. Extrapolations from only one of the 160\,$\mu$m, 250\,$\mu$m or 350\,$\mu$m bands alone tend to overestimate the total IR luminosity. This may be explained by the lack of far-IR constraints around and above $\sim$150\,$\mu$m (rest-frame) before Herschel on those templates. We also note that the dust temperature of luminous IR galaxies (LIRGs, L$_{\rm IR}$$\geq$10$^{11}$ L$_{\sun}$) around $z$$\sim$1 is mildly colder by 10-15\,\% than their local analogs and up to 20\,\% for ULIRGs at $z$$\sim$1.6 (using a single modified blackbody-fit to the peak far-IR emission with an emissivity index of $\beta$=1.5). Above $z$=1.5, distant galaxies are found to exhibit a substantially larger mid- over far-IR ratio, which could either result from stronger broad emission lines or warm dust continuum heated by a hidden AGN. Two thirds of the AGNs identified in the field with a measured redshift exhibit the same behavior as purely star-forming galaxies. Hence a large fraction of AGNs harbor coeval star formation at very high SFR and in conditions similar to purely star-forming galaxies. 

}
\keywords{Galaxies: evolution -- Galaxies: active -- Galaxies: starburst -- Infrared: galaxies
               }

\maketitle

%________________________________________________________________
\section{Introduction}
%________________________________________________________________
The mechanisms that govern star formation in galaxies are poorly understood: recent evidence points to relatively steady--state growth rather than episodic, merger--driven starbursts, with a tight link between galaxy mass and star-formation rate (SFR; Noeske et al. 2007, Elbaz et al. 2007, Daddi et al. 2007). To a large degree, these uncertainties arise because observations to date have only been sensitive to a small fraction of the bolometric energy emerging from dusty star formation. At high redshift, most of the energy from star formation (SF) and active galactic nuclei (AGN) is absorbed by dust (and gas) and re-radiated at infrared wavelengths. ISO and Spitzer studies have suggested that luminous IR galaxies (LIRGs, 10$^{12}$$>$L$_{\rm IR}$/L$_{\sun}$$\geq$10$^{11}$) dominate global SF at $z$$\sim$1 (Chary \& Elbaz 2001 - hereafter CE01, Le Floch et al. 2005, Magnelli et al. 2009), while submm and Spitzer data demonstrate that ultra-luminous IR galaxies (ULIRGs, L$_{\rm IR}$$\geq$10$^{12}$ L$_{\sun}$) are equally important at $z$$\sim$2 (Papovich et al. 2007, Caputi et al. 2007, Daddi et al. 2007, Magnelli et al. 2009, 2010). Until the launch of the {\it Herschel} Space Observatory (Pilbratt et al. 2010), these analyses relied strongly on substantial extrapolation from the mid-IR or sub-mm or on even more uncertain corrections of the UV luminosity (Buat et al. 2009, Daddi et al. 2007). Our aim is to determine how accurate these extrapolations are and search for evidence of a major change in the IR properties of galaxies with increasing redshift, using the combined power of the PACS (Poglitsch et al. 2010) and SPIRE (Griffin et al. 2010) instruments. Due to the effects of k-correction and sensitivity limitations from source confusion, few individual galaxies are detected in all {\it Herschel} bands, hence the need to determine the robustness of extrapolations of total IR luminosities from one or more bands in the mid or far IR.

We will use below a cosmology with $H_0$=70~$\mathrm{km s^{-1} Mpc^{-1}}$, $\Omega_M=0.3$, $\Omega_\Lambda=0.7$.

%________________________________________________________________
\section{Sample and method}
\label{SEC:method}
%________________________________________________________________
%sample
The sample consists of galaxies observed in the GOODS-North field within the PEP\footnote{www.mpe.mpg.de/ir/Research/PEP} and HerMES\footnote{hermes.sussex.ac.uk} (Oliver et al., in prep.) guaranteed time key programs. Measurements of the PACS-100, 160\,$\mu$m (PEP) and SPIRE-250, 350, 500\,$\mu$m (HerMES) flux densities and S$/$N ratios were obtained from point source fitting on 24\,$\mu$m priors by S.Berta, B.Magnelli (PEP, Berta et al. 2010, Magnelli et al. 2010) and I.Roseboom (HerMES, Roseboom et al., in prep.) respectively. 

Within a field of 10\arcmin$\times$15\arcmin, which has coverage with Spitzer and HST--ACS, there are a total of 1468 24\,$\mu$m sources which meet the reliability criterion of S$/$N$\geq$5 and F$_{24}$$\geq$20\,$\mu$Jy (see Magnelli et al. 2009). 95.6\,\% of them either have a spectroscopic (67.2\,\%, Cohen et al. 2000, Wirth et al. 2004, Barger et al. 2008) or photometric (28.3\,\%, from Le Borgne et al. 2009) redshift. {\it Herschel} detects a third of them (493 galaxies) in at least one of the PACS or SPIRE passbands. PACS measurements are used down to the 3-$\sigma$ limits of 3mJy and 5.7mJy at 100\,$\mu$m and 160\,$\mu$m respectively. Both PACS passbands lie above the confusion limit (Berta et al. 2010).
We use SPIRE measurements down to the 5-$\sigma$ limits of the prior catalog of 4.4mJy, 4.8mJy and 7.6mJy at 250\,$\mu$m, 350\,$\mu$m and 500\,$\mu$m respectively. We note that these measurements lie below the SPIRE confusion limit of 5.8mJy, 6.3mJy, 6.8mJy 1-$\sigma$ (Nguyen et al. 2010). However, this limit is a spatially averaged statistical limit which considers that galaxies are homogeneously distributed in the field and are all affected in the same way by close neighbors. Here we take advantage of the higher spatial resolution at lower wavelengths (24\,$\mu$m and PACS bands) to flag galaxies more "isolated" than others for which SPIRE flux densities can potentially be more robust. To do this, we require at least one detection in one of the PACS bands for all galaxies studied here and flag as "clean" galaxies those sources which have at most one bright neighbor within 20$\arcsec$ (close to the full width half maximum, FWHM, of the central {\it Herschel} wavelength of 250\,$\mu$m) with F$_{24}$$>$50\% of the central 24\,$\mu$m source.

Below, all IR luminosities are extrapolated to the 8--1000\,$\mu$m range, i.e. $L_{\rm IR}^{\lambda}$ is the total IR luminosity estimated from only one passband at $\lambda$. The reference total IR luminosity, $L_{\rm IR}^{\rm tot}$[Herschel], was determined from the best fit of at least two photometric measurements above 30\,$\mu$m (rest-frame), using the whole library of SED templates from CE01 independently of their luminosity. We did the same computation with the full Dale \& Helou (2002, DH02) library of template SEDs and found that both $L_{\rm IR}^{\rm tot}$ agreed within 12\,\%, with a median ratio of 1. Error bars on $L_{\rm IR}^{\rm tot}$ were obtained by randomly selecting flux densities in the {\it Herschel} bands within their associated error bars and fitting them again. A total of 222 galaxies have two photometric measurements above 30\,$\mu$m (rest-frame), among which 140 are "clean" star-forming galaxies, 55 are "non-clean" and 27 show the signature of an AGN. AGNs were identified from either L$_{\rm X}$[0.5-8.0 keV] $>$ 3$\times$10$^{42}$ ergs s$^{-1}$, a hardness ratio (ratio of the counts in the 2-8 keV to 0.5-2 keV passbands) higher than 0.8, N$_{\rm H}\geq$10$^{22}$ cm$^{-2}$, or broad / high-ionization AGN emission lines (Bauer et al. 2004). We then computed total IR luminosities from a single passband, $L_{\rm IR}^{\lambda}$, using the CE01 technique, i.e. a fixed SED was attributed to every $L_{\rm IR}$ and chosen to best fit the observed 24\,$\mu$m measurement at the redshift of the source ($L_{\rm IR}^{24\,\mu{\rm m}}$). We used the same technique to extrapolate $L_{\rm IR}^{\lambda}$ for each one of the PACS or SPIRE passbands at 100\,$\mu$m, 160\,$\mu$m, 250\,$\mu$m and 350\,$\mu$m. Only 10 galaxies with a redshift were detected in one of the PACS bands and at 500\,$\mu$m. For comparison, we computed $L_{\rm IR}^{\lambda}$ with the libraries of DH02 and Lagache et al. (2004, LDP04). For DH02, a single SED was attributed to a given $L_{\rm IR}^{\rm tot}$, after correlating the luminosity to the 60/100\,$\mu$m ratio (see Marcillac et al. 2006). 

%stacking
The analysis was extended below the PACS detection limits by stacking the PACS-160\,$\mu$m images at the positions of 24\,$\mu$m sources (on residual images after PSF-subtracting the PACS sources above the 3-$\sigma$ limit).
%dust temperatures
Dust temperatures were computed as in Hwang et al. (in prep.), i.e. we used a modified blackbody fit with a $\beta$=1.5 emissivity index to the galaxies for which {\it Herschel} measurements exist on both sides of the peak far-IR emission. Error bars on T$_{\rm dust}$ were derived in the same way as for $L_{\rm IR}^{\rm Herschel}$.
%________________________________________________________________
\section{Results and discussion}
\label{SEC:results}
\begin{figure}%*}[ht!]
   \centering
   \includegraphics*[width=8.5cm]{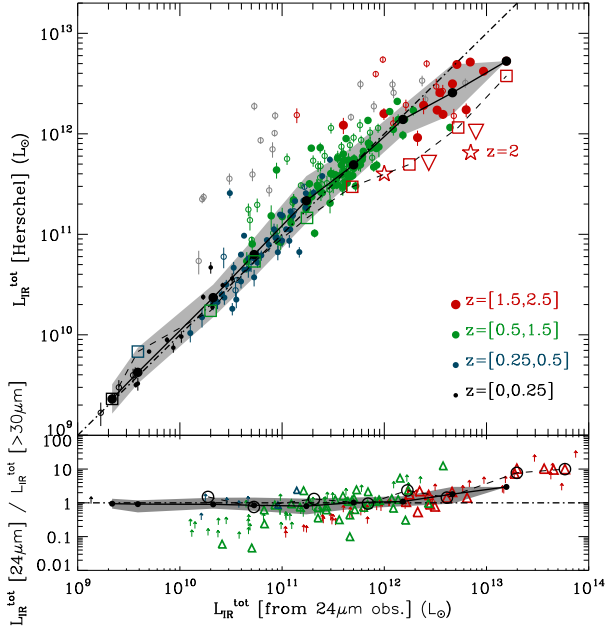}
      \caption{$L_{\rm IR}^{\rm tot}$[Herschel] versus $L_{\rm IR}^{\rm tot}$[24\,$\mu$m] for galaxies with at least 2 {\it Herschel} detections and no AGN signature (filled dots in upper panel) and AGNs detected at 24\,$\mu$m and in at least one {\it Herschel} passband (open triangles in lower panel). Both $L_{\rm IR}$ cover 8--1000\,$\mu$m. {\bf (Upper panel)} "clean" (filled dots) and "non-clean" (open circles: grey, no PACS detection; black, 1 PACS detection) galaxies. Plain line, black dots and grey zone: median of clean sample and 68\,\% envelope ($\sim$0.15 dex rms). Stacking measurements per 24\,$\mu$m luminosity bins: present study (open squares), Nordon et al. (2010) for $z=1.5-2.5$ galaxies (orange triangles). Orange open stars: Daddi et al. (2007, based on UV corrected for extinction). {\bf (Lower panel)} 24\,$\mu$m AGNs detected in one {\it Herschel} band (open triangles) or with {\it Herschel} upper limits (vertical arrows). Black open circles and dashed line: stacked values combined with detections weighted with source numbers. }
         \label{FIG:complir}
   \end{figure}

%________________________________________________________________
The direct comparison of the 8--1000\,$\mu$m luminosities derived from the 24\,$\mu$m band alone ($L_{\rm IR}^{\rm tot}$[from 24\,$\mu$m]) and from {\it Herschel} above 30\,$\mu$m (rest-frame, $L_{\rm IR}^{\rm tot}$[Herschel]) shows a remarkable consistency over a redshift range of $z$=0--1.5 and over three decades in luminosity up to ULIRGs (Fig.~\ref{FIG:complir}). The median trend (filled black circles and plain line) remains within 10\,\% of the one-to-one correlation and the envelope including 68\,\% of the galaxies above and below the median (grey zone) has a width of $\pm$0.15 dex (40\,\%), which is similar to the dispersion measured locally (see CE01). Note a cloud of galaxies with an excess far-IR emission mostly composed of objects that we flagged as "non-clean", i.e. with close bright neighbors (open symbols). The largest excess is found for galaxies with SPIRE measurements only (and at least two SPIRE fluxes, light grey open circles), suggesting that their SPIRE fluxes are boosted by close neighbors. Very few "clean" galaxies lie outside the 0.15 dex dispersion, which suggests that SPIRE flux densities can be trusted even at very faint levels, i.e. below the statistical confusion limit. The tight correlation remains even after combining detections with stacked measurements in bins of 24\,$\mu$m derived luminosities (large open squares), except above $z$$=$1.5 (red symbols).
%-------------------------------------------------------------
   %*}
%-------------------------------------------------------------

The tight correlation between $L_{\rm IR}^{\rm tot}$ as derived from 24\,$\mu$m and {\it Herschel} below $z$=1.5, which extends over three decades in luminosity, is puzzling because galaxies have strongly evolved over the last 9 billion years (70\,\% of the Universe age) during which most present-day stars formed (gas mass fraction, metallicity, compactness, dynamical status, e.g. mergers). It is as puzzling to see that AGNs follow a similar trend as star-forming galaxies (open triangles in lower panel of Fig.~\ref{FIG:complir}). We note that among the AGNs with a redshift (either spectro- or photo-metric for 83\,\% of all AGNs in the field), 70\,\% are detected at 24\,$\mu$m (arrows) and 31\,\% in at least one the {\it Herschel} bands (open triangles). While it may be understood that at low redshifts, hence at large wavelengths, the warm dust continuum heated by the AGN remains negligible at 24\,$\mu$m (quasar spectra drop beyond $\sim$20\,$\mu$m which Netzer et al. 2007 interpret as the signature of a minimum temperature of $\sim$200 K), the star-formation activity in AGNs could have been associated to more compact geometries, e.g. due to a merger, and presented different IR signatures. But whether IR galaxies harbor an AGN or not, their star-formation activity produces a similar radiation pattern suggesting no major differences in the coeval activity of star formation with respect to purely star-forming galaxies. The combination of stacking and detections (weighted by number of sources, large open circles in Fig.~\ref{FIG:complir}-bottom) confirms the same trend as for the star-forming galaxies. 

%-------------------------------------------------------------
   \begin{figure}%[ht!]
   \centering
   \includegraphics*[width=8cm]{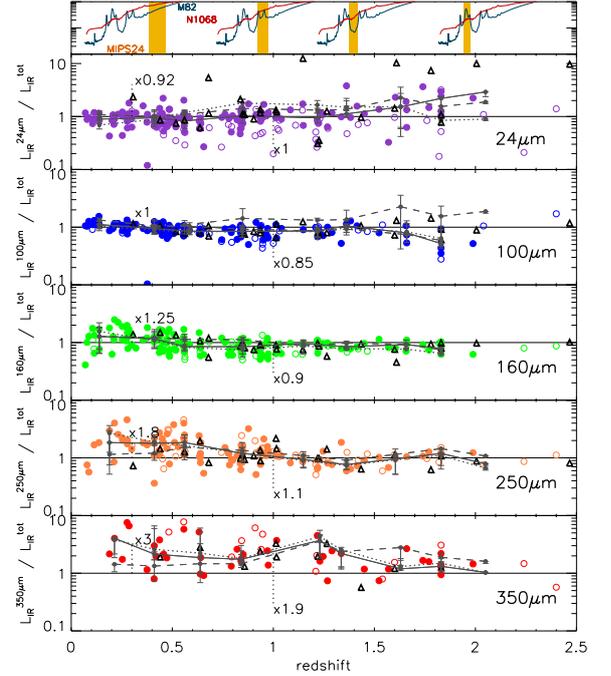}
      \caption{Redshift evolution of $L_{\rm IR}$ derived from monochromatic mesurements at 24\,$\mu$m, 100\,$\mu$m, 160\,$\mu$m, 250\,$\mu$m and 350\,$\mu$m over  $L_{\rm IR}^{\rm Herschel}$[$>$30\,$\mu$m] (filled dots: clean, open dots: non clean  galaxies). Only 10 "clean" galaxies have a 500\,$\mu$m measurement. Black triangles: AGNs. Lines: median ratios for local SEDs (CE01: plain, DH02: dashed, LDP04: dotted line). Upper panel: SEDs of M82 (starburst) and NGC 1068 (type II AGN) and MIPS 24\,$\mu$m filter at $z$=0.25,0.9,1.4,2.
      }
         \label{FIG:monolum}
   \end{figure}
%-------------------------------------------------------------
Above $z$=1.5, {\it Herschel} confirms that the mid-IR overestimates $L_{\rm IR}^{\rm tot}$ by factors of about 2--3 for the detected ULIRGs, but up to 5--7 when stacking is combined to detections (see Nordon et al. 2010, orange triangles in upper panel of Fig.~\ref{FIG:complir}). This confirms the claim from previous works based on Spitzer 70\,$\mu$m stacking or UV corrected for extinction (Daddi et al. 2007, Papovich et al. 2007, Magnelli et al. 2010), but also provides a way to revise local IR SEDs accordingly. The upper panel of Fig.~\ref{FIG:monolum} shows the region of the mid-IR SED that is sampled by the 24\,$\mu$m Spitzer passband as a function of redshift. It is clear that the mid-IR excess luminosities are found when the filter is centered on the 7.7\,$\mu$m PAH complex. The discrepancy could either be explained by a larger PAH emission in distant ULIRGs or instead by the hot dust heated by a buried AGN. We note however, that part of this discrepancy could be due to uncertain local SEDs templates (Takeuchi et al. 2005, Buat et al. 2009). In Fig.~\ref{FIG:monolum}, we follow $L_{\rm IR}^{\lambda}$ over $L_{\rm IR}^{\rm tot}$ for each {\it Herschel} band and for the 24\,$\mu$m one. Again, AGNs (black open triangles) follow the same pattern as purely star-forming galaxies (which may include heavily obscured AGNs) and "clean" galaxies (filled circles) are less dispersed than "non-clean" ones (open circles). The points in Fig.~\ref{FIG:monolum} were computed using the CE01 SEDs. In order to see the effect of using SEDs from DH02 or LDP04, we materialized their median ratio and 68\,\% dispersion by dashed (DH02) and dotted (LDP04) lines with error bars.

\begin{figure}%[ht!]
   \centering
   \includegraphics*[width=8cm]{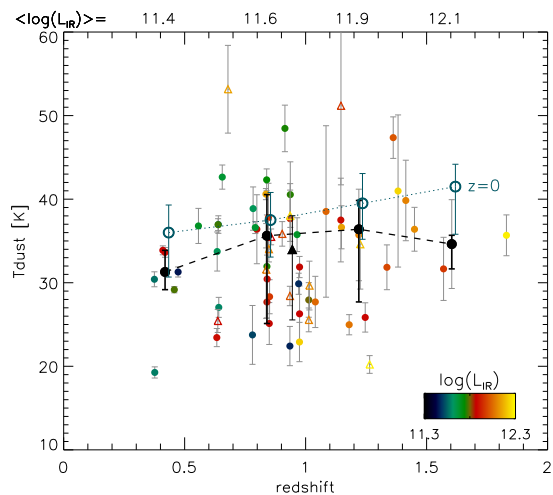}
      \caption{T$_{\rm dust}$ versus redshift for galaxies with 11.3$\leq$log(L$_{\rm IR}$/L$_{\sun}$)$\leq$12.3. Filled circles: "clean" galaxies, median in black with 68\,\% envelope. Triangles: AGNs, median in black. Error bars on T$_{\rm dust}$ from Monte Carlo in Herschel error bars (see Hwang et al., in prep.). Empty circles: median T$_{\rm dust}$ of local galaxies (computed with the same technique i.e. MBB with $\beta$=1.5, see Hwang et al., in prep.) for the median log(L$_{\rm IR}$) in each z bin (values in upper x-axis) again with associated 68\,\% envelope.}
         \label{FIG:evoltdust}
   \end{figure}

We note that the CE01 templates provide the best $L_{\rm IR}^{\rm tot}$ from 24\,$\mu$m, 100\,$\mu$m and 160\,$\mu$m while DH02 is better in the SPIRE bands. However, monochromatic derivations of IR luminosities from the 160\,$\mu$m, 250\,$\mu$m or 350\,$\mu$m values alone tend to overestimate the true $L_{\rm IR}^{\rm tot}$ with all three libraries of template SEDs below $z$$\sim$1--1.5 (e.g. factors of 1.25, 1.8 and 3 with CE01 at $z$$\sim$0.3). This enhanced emission above 150\,$\mu$m with respect to existing templates suggests a colder dust temperature than previously inferred due to the lack of constraints before {\it Herschel} at these far-IR wavelengths (see below and Rowan-Robinson, in prep.). To assess the evolution of far-IR color temperatures with redshift, we selected all galaxies within a decade in luminosity (L$_{\rm IR}$=10$^{11.3}$--10$^{12.3}$ L$_{\sun}$), i.e. $\sim$LIRGs, which span a redshift range of 0.5$<$$z$$<$1.5 (Fig.~\ref{FIG:evoltdust}). T$_{\rm dust}$ was measured assuming a single modified blackbody-fit to their peak far-IR emission with an emissivity index of $\beta$=1.5. We found that the average color temperature is $\sim$35 K and within the dispersion does not show any strong evolution with redshift but appears to be systematically colder than a sample of local galaxies of similar median luminosities (Hwang et al., in prep.) by $\sim$10--15\,\% and up to $\sim$20\,\% (8K)  at $z$$\sim$1.6 for $L_{\rm IR}^{\rm tot}$$\sim$10$^{12}$ L$_{\odot}$.
%-------------------------------------------------------------
%-------------------------------------------------------------

We also note that AGNs (open triangles, and median in filled black triangle) exhibit similar T$_{\rm dust}$ than purely star-forming galaxies. This again shows that in the regime where star formation dominates the IR emission of AGNs, this activity does not differ from normal star-forming galaxies (see also Shao et al 2010, Hatziminaoglou et al. 2010).

\section{Conclusion}
We show that measurements in the SPIRE bands can be boosted by bright neighboring galaxies due to the large beam size. However, we were able to flag "clean" sources in more isolated environments for which fluxes below the statistical 5-$\sigma$ confusion limit appear to be robust. This may also apply to submm measurements with ground-based telescopes whose PSF FWHM is similar. Using such a "clean" sample of galaxies we were able to obtain the following results.

Below $z$$\sim$1.5, extrapolations of the total IR luminosity of galaxies based on mid-IR measurements agree with those measured directly with {\it Herschel}
over three decades in luminosity below the ULIRG regime, with a dispersion of only 40\,\% (0.15 dex), and are therefore similar to the local one. This narrow distribution is puzzling when considering the range of physical processes (gas mass fraction, different geometries, grain size distribution, metallicity) that could have affected the SED of these galaxies which dominated the SFR density of the Universe over 80\,\% of the age of the Universe.

When used alone, each of the far-IR bands at 160\,$\mu$m, 250\,$\mu$m and 350\,$\mu$m tend to overpredict the total IR luminosity by factors of 1.25, 1.8 and 3 respectively with e.g. the CE01 templates, due to the lack of constraints on those templates at far IR wavalengths around and above $\lambda$$\sim$150\,$\mu$m prior to {\it Herschel}. We also note that the dust temperature of $z$=1-1.5 LIRGs -- estimated from a modified blackbody-fit with an emissivity index of $\beta$=1.5 -- is $\sim$35 K, hence about 10-15\,\% colder than their local analogs and 20\,\% colder for ULIRGs at $z$$\sim$1.6.

At $z$$>$1.5, IR SEDs are substantially different from local ones as suggested by previous studies (Daddi et al. 2007, Papovich et al. 2007, Magnelli et al. 2010). L$_{\rm IR}$ derived from 24\,$\mu$m alone is overestimated by a factor 2--3 in the ULIRGs detected with {\it Herschel} and up to $\sim$7 from stacking (Nordon et al. 2010). 

About a third of the AGNs with a redshift are detected with {\it Herschel} and an extra 40\,\% at 24\,$\mu$m alone, allowing us to extend our study from stacking to 70\,\% of the AGNs. We find that below $z$$\sim$1.5, AGNs exhibit a similar L$_{\rm IR}^{24}$/L$_{\rm IR}^{\rm Herschel}$ ratio than star-forming galaxies of similar L$_{\rm IR}$ and that their dust temperature is similar to that of purely star-forming galaxies. This suggests that their IR emission is dominated by star formation, hence that SF and AGN activity happen concomitantly in these galaxies. This result confirms that a large fraction of AGNs do harbor intense star formation activity of several 10 or 100 solar masses per year in conditions similar to purely star-forming galaxies (see also Shao et al 2010, Hatziminaoglou et al. 2010). 

Through deeper observations down to the 100\,$\mu$m confusion limit with the GOODS-{\it Herschel} key program (P.I.: D. Elbaz), we will be able to study the SED of galaxies to fainter luminosities and higher redshifts than undertaken in this paper.
%__________________________________________________________________

\begin{acknowledgements}
PACS has been developed by a consortium of institutes led by MPE (Germany) and including UVIE
(Austria); KU Leuven, CSL, IMEC (Belgium); CEA, LAM (France); MPIA (Germany); INAFIFSI/
OAA/OAP/OAT, LENS, SISSA (Italy); IAC (Spain). This development has been supported by the
funding agencies BMVIT (Austria), ESA-PRODEX (Belgium), CEA/CNES (France), DLR (Germany),
ASI/INAF (Italy), and CICYT/MCYT (Spain).
SPIRE has been developed by a consortium of institutes led by Cardiff University (UK) and including Univ.
Lethbridge (Canada); NAOC (China); CEA, LAM (France); IFSI, Univ. Padua (Italy); IAC (Spain);
Stockholm Observatory (Sweden); Imperial College London, RAL, UCL-MSSL, UKATC, Univ. Sussex
(UK); and Caltech, JPL, NHSC, Univ. Colorado (USA). This development has been supported by national
funding agencies: CSA (Canada); NAOC (China); CEA, CNES, CNRS (France); ASI (Italy); MCINN
(Spain); Stockholm Observatory (Sweden); STFC (UK); and NASA (USA).
\end{acknowledgements}

\end{document}